\documentclass[pra,twocolumn,showpacs,floatfix]{revtex4}
\usepackage{graphicx}
\begin{document}

\title{Rotational properties of dipolar Bose-Einstein condensates 
confined in anisotropic harmonic potentials}
\author{F. Malet$^1$, T. Kristensen$^2$, S. M. Reimann$^1$, and 
G. M. Kavoulakis$^3$}
\affiliation{$^1$Mathematical Physics, Lund Institute of Technology,
P.O. Box 118, SE-22100 Lund, Sweden \\
$^2$\'Ecole Normale Sup\'erieure de Cachan, Cachan cedex, 94230 Cachan, 
France \\
$^{3}$Technological Educational Institute of Crete, P.O. Box 1939,
GR-71004, Heraklion, Greece}

\date{\today}

\begin{abstract}

We study the rotational properties of a dipolar Bose-Einstein 
condensate confined in a quasi-two-dimensional anisotropic 
trap, for an arbitrary orientation of the dipoles with respect 
to their plane of motion. Within the mean-field approximation
we find that the lowest-energy state of the system depends strongly 
on the relative strength between the dipolar and the contact 
interactions, as well as on the size and the orientation of the 
dipoles, and the size and the orientation of the deformation of 
the trapping potential. 

\end{abstract}


\maketitle

\section{Introduction}

The rotational properties of Bose-Einstein condensates 
have been studied extensively over the last decades, 
beginning with the well-known experiments performed on 
liquid Helium. Bose-Einstein-condensed gases of ultra-cold
atoms have provided us with an ideal system for the 
study of this problem \cite{Leggett,AF,Vortices}: 
In addition to the fact that they are dilute, many 
of the system parameters are tunable externally, 
making it possible to explore experimentally many 
novel properties. So far most of the experiments 
have been performed in isotropic \cite{Mad00,m2,m3,m4,m5,m6,m7,m8}, 
as well as anisotropic \cite{Hod01} harmonic potentials. More 
recently, other trapping potentials such as, for example, 
anharmonic \cite{Bretin} and toroidal \cite{BP} ones, have 
also been addressed.

While the initial experiments considered atoms with 
interactions that could be modelled via an isotropic 
and local, i.e., contact, potential, more recently 
the behavior of atoms with an electric or 
magnetic dipole moment has been investigated 
\cite{Dipteo1,Dipteo2}. The physics of these gases 
is rather different because the dipolar interaction 
is anisotropic, non-local, and finally it can be both 
attractive -- when the dipoles are placed head-to-tail 
-- and repulsive -- when they are placed side-to-side.
Initially dipolar effects were investigated experimentally 
in $^{52}$Cr atoms \cite{Gri05}. More recently dipolar 
effects have also been investigated experimentally in 
condensates of $^{39}$K and $^7$Li \cite{Fat08} atoms, 
in spinor condensates \cite{Ven08}, in ultracold gases 
of Dy atoms \cite{Lu10}, and in ultracold polar molecules 
\cite{Osp06,Osp08}. These studies have focused on different 
aspects of the dipolar interaction such as, e.g., its 
anisotropy \cite{Fat08}, or on its attractive character 
and the consequent possible collapse of the gas \cite{Koch08}.
 
The rotational properties of Bose-Einstein condensates are
described e.g. in the review articles \cite{Leggett,AF,Vortices}, 
as well as in Refs.\,\cite{SFetter,Rok99,Fet01,Okt04,Shl,Aft09,Fet10}. 
In particular, anisotropic trapping potentials have been 
considered in Refs.\,\cite{Fet01,Okt04,Shl,Aft09,Fet10}. 
Dipolar gases have also been extensively addressed theoretically, 
see for example 
Refs.\,\cite{Dipteo1,Dipteo2,San00,Gor00,Gio02,Yi06,ODell07,Kom07,Abad09,Cai10,Cre10},
and studies on dipolar Bose-Einstein condensates under rotation
have shown that the vortex properties of these systems are strongly affected 
by the anisotropic character of the dipole-dipole interaction 
and by the relative strength between the dipolar and the contact
interactions \cite{Yi06,Kom07,Abad09,Bij07,Sim11}. 
The relative orientation of the 
dipoles with respect to the confining potential also affects 
the critical rotation frequency for the nucleation of a 
vortex state in the condensate, lowering (raising) its value 
with respect to the non-dipolar case when the trap is 
elongated in the direction perpendicular (parallel) to the 
orientation of the dipoles \cite{ODell07}.

Some of the above studies have employed a fully three-dimensional 
approach to solve the Gross-Pitaevskii equation \cite{ODell07,Abad09}, 
whereas other works have examined quasi-one- or quasi-two-dimensional 
systems \cite{Yi06,Kom07,Cai10,Cre10}. In most of these papers, 
the orientation of the dipoles has been limited to only two 
possibilities: perpendicular, or parallel to their plane of motion. 
Very recently, the more general and interesting case of an arbitrary 
orientation of the dipole moment of the atoms has been investigated 
for non-rotating gases \cite{Cai10,Sap10,Cre10}.

Here, we study the rotational properties of a quasi-two-dimensional 
dipolar Bose-Einstein condensate confined in a rotating anisotropic 
harmonic trapping potential, with an arbitrary orientation
of the dipoles. Since the latter can be controlled by means
of external electric or magnetic fields, this is also
very relevant from the experimental point of view.
As compared to the case where the gas is 
trapped in an axially-symmetric trap and interacts with a 
contact potential, here the symmetry is broken in two separate 
and independent ways: Through the anisotropy of the trap, and 
also through the directional dependence of the dipolar interaction. 
The combination of these two symmetry-breaking mechanisms 
gives rise to interesting vortex configurations, which are 
investigated in this article. 

In what follows we first present our model in Sec.\,II. 
In Sec.\,III we investigate the rotational properties of the 
system as a function of the trap anisotropy, the strength of 
the dipolar interaction, and the orientation of the dipole 
moment of the atoms with respect to their plane of motion. 
Finally, in Sec.\,IV we summarize our results and conclude.

\section{Model and methodology}

Let us consider a Bose-Einstein condensate of atoms of mass 
$M$ confined by some external potential, which is harmonic 
in all directions,
\begin{eqnarray}
V_T({\bf r}) = \frac 1 2 M \omega_z^2 z^2 + V({\bf r}_{\bot}).
\end{eqnarray}
Here $\omega_z$ is the frequency of the potential along the 
$z$ axis, which we assume to be very tight, while in the 
$xy$ plane the potential is taken to be anisotropic,
\begin{eqnarray}
V({\bf r}_{\bot})
&=&\frac{1}{2}M\omega_0^2(x^2+y^2)+2 A M\omega_0^2(x^2-y^2) 
\nonumber\\
&\equiv&\frac{1}{2}M(\omega_x^2x^2+\omega_y^2y^2).
\end{eqnarray}
In the above expression ${\bf r_{\bot}} = (x,y)$, and $\omega_x 
\equiv \omega_0 \sqrt{1+4A}$, $\omega_y \equiv \omega_0 \sqrt{1-4A}$ 
are the trap frequencies along the $x$ and $y$ axes, 
respectively. Here $A$ is a dimensionless parameter 
determining both the direction and the strength of the 
deformation of the trap.

The assumption of a very tight potential along the $z$ 
axis, $\hbar \omega_z$ being much larger than any other
energy scale in the problem, allows us to treat the 
problem as quasi-two-dimensional. With this assumption, 
the degrees of freedom along the $z$ axis are frozen and 
the system occupies only the lowest-energy eigenstate 
$\phi_0(z)$ of the potential $V(z) = M \omega_z^2 z^2/2$. 
As a result, the order parameter $\Psi({\bf r})$ separates 
to $\Psi({\bf r}) = \psi({\bf r}_{\bot}) \phi_0(z)$.
\begin{figure}[t]
\centerline{\includegraphics[width=7cm,clip]{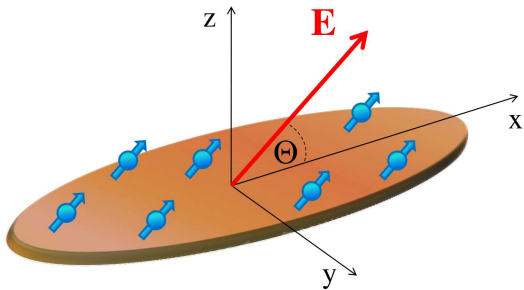}}
\caption{(Color online)
Schematic illustration of the quasi-two-dimensional elliptical
confining potential. A very tight potential is assumed along 
the $z$ axis (not shown in the plot). The atoms have a dipole 
moment in the $xz$ plane, at some direction that forms an angle 
$\Theta$ with the $x$ axis, due to the action of an external 
electric (or magnetic) field ${\bf E}$.} 
\label{fig1}
\end{figure}
We also assume that the atoms have a non-zero (electric or 
magnetic) dipole moment, and that all the dipoles are aligned 
to some external (electric or magnetic) field, which is in 
the $xz$ plane, and forms an angle $\Theta$ with the $x$ axis, 
as illustrated schematically in Fig.\,\ref{fig1}. In three 
dimensions, the dipolar interaction between two atoms separated 
by the vector ${\bf r}$ is given by \cite{Dipteo2}
\begin{equation}
V_{dd}(\mathbf{r})=D^2\frac{1-3\cos^{2}\theta_{rd}}{r^{3}}.
\label{Vdd}
\end{equation}
In the above expression, $D^2 = d^2/(4 \pi \epsilon_0)$
when the atoms have an electric dipole moment $d$, 
where $\epsilon_0$ is the permittivity of the vacuum. When
the atoms have a magnetic moment $\mu$, $D^2 = \mu_0 \mu^2/(4 \pi)$, 
where $\mu_0$ is the permeability of the vacuum. 
Also, $\theta_{rd}$ is the angle between the 
polarization direction and the relative position vector 
${\bf r}$ of the two dipoles.

As shown in Refs.\,\cite{Cai10,Cre10}, in order to derive an 
effective potential for the two-dimensional problem, one 
assumes the Gaussian profile of the harmonic oscillator 
in the $z$ direction $\phi_0(z)$. Then, integrating 
$V_{dd}(\mathbf{r})$ over the $z$ coordinates, 
\begin{eqnarray}
   V_{\rm eff}({\bf r}_{\bot}) = \int
|\phi_0(z)|^2 V_{dd}(\mathbf{r}-\mathbf{r}') 
|\phi_0(z')|^2 \, dz dz'.
\label{intz}
\end{eqnarray}
Working with cylindrical polar coordinates 
${\bf r}_{\bot} = (r_{\bot},\phi)$,
\begin{eqnarray}
   V_{\rm eff}({\bf r}_{\bot})=
  \frac{D^{2}}{\sqrt{8 \pi}}
 \frac{e^{w/2}}{a_z^3}\biggl\{(2+2w)K_{0}(w/2)-2w
 K_{1}(w/2)
\nonumber\\
+\cos^{2}\Theta
\biggl[-(3+2w)K_{0}(w/2)+(1+2w)K_{1}(w/2)\biggl]
\nonumber\\
+2\cos^{2}\Theta\cos^{2}\phi\biggl[-w
K_{0}(w/2)+(w-1)K_{1}(w/2)\biggl]\biggl\}.
\label{eqVdipeff}
\end{eqnarray}
Here $a_z=\sqrt{\hbar/m\omega_z}$ is the oscillator length 
in the $z$ direction, $w \equiv r_{\bot}^2/2a_z^2$, and 
$K_0(w)$ and $K_1(w)$ are the zero-order and first-order modified 
Bessel functions of the second kind. It can be checked easily 
that for the particular cases $\Theta = 0^\circ$ and $90^\circ$, 
Eq.\,(\ref{eqVdipeff}) reduces to the expressions found 
in Refs.\,\cite{Yi06,Kom07}. The assumed quasi-two-dimensional 
behavior of the system allows us to derive the above expression 
for the dipolar interaction, and also allows us to consider any 
arbitrary orientation of the dipoles.

In addition to the dipolar interaction, we also consider 
the usual contact potential between the atoms, given 
by $V_{\rm int}({\bf r}) = U_0 \delta ({\bf r})$, where 
$U_0 = 4 \pi \hbar^2 a/M$. Here $a$ is the scattering 
length for zero-energy elastic atom-atom collisions. 
Again, integrating over the profile along the $z$ axis, 
the corresponding effective two-dimensional contact potential
becomes $V_s({\bf r}_{\bot}) = g \delta({\bf r}_{\bot})$, 
where $g$ is given by $g = U_0 \int |\phi_0(z)|^4 dz = 
U_0/(\sqrt{2 \pi} a_z)$.

Assuming also that the trap rotates around the $z$ axis with 
some angular frequency $\Omega$, the Gross-Pitaevskii equation 
for the order parameter $\psi({\bf r}_{\bot})$ takes the 
following, nonlocal form
\begin{eqnarray}
\biggl[- \frac {\hbar^2 \nabla_{\bot}^2}{2 M} + V({\bf r}_{\bot}) 
+ V_{\rm dip}({\bf r}_{\bot}) + g |{\psi({\bf r}_{\bot})}|^2  
&-&\Omega L_z\biggl]\psi({\bf r}_{\bot}) 
\nonumber \\
&=& \mu \psi({\bf r}_{\bot}),
\label{gpe}
\end{eqnarray}
where $L_z$ is the operator of the angular momentum along the $z$ 
axis, $\mu$ is the chemical potential, and
\begin{equation}
  V_{\rm dip}({\bf r}_{\bot}) = \int 
 V_{\rm eff}({\bf r}_{\bot}-{\bf r}_{\bot}^{\prime}) \,
\left|\psi({\bf r}_{\bot}^{\prime})\right|^2\,d{\bf r}_{\bot}^{\prime}
\end{equation}
is the dipolar interaction potential. We have treated this term making use 
of the convolution theorem and fast-Fourier-transform techniques, 
combined with the introduction of a cutoff at small distances, 
where $V_{\rm eff}({\bf r}_{\bot})$ diverges, as done in 
Ref.\,\cite{Gor00}.

In order to solve Eq.\,(\ref{gpe}) we have employed a fourth-order 
split-step Fourier method within an imaginary-time propagation 
approach \cite{Chi05}. Thus, starting with a reasonable initial 
state we propagate it in imaginary time until we reach a steady 
state, after a sufficiently large number of time steps. 

\section{Results -- Rotational properties of a driven gas}

Since there is a large number of parameters in the present
problem, we fix $\omega_z$ and $\omega_0$ throughout this 
study, choosing $\omega_z/\omega_0 = 100$. 

One useful dimensionless quantity is the ratio between the 
interaction energy per particle due to the contact potential 
$E_s$ and the oscillator energy $\hbar \omega_0$. For a cloud 
of homogeneous density of radius equal to the oscillator 
length $a_{0} = \sqrt{\hbar/M \omega_0}$ the interaction 
energy is $E_s = g N/(\pi a_{0}^2)$, where $N$ is the 
atom number. The ratio $E_s/\hbar \omega_0$ is thus equal
to $gNM/(\pi \hbar^2)$, and in what follows we choose it 
equal to $100/\pi$. 

Another dimensionless quantity that is convenient to 
introduce is the ratio between the ``dipolar length''
$a_{dd}$, and the $s-$wave scattering length $a$, which 
we denote as $\varepsilon_{dd} \equiv a_{dd}/a$ 
\cite{Dipteo2}. Defining $a_{dd}\equiv M D^2/(3 
\hbar^2)$, then $\varepsilon_{dd} = 4 \pi D^2/3 U_0$. 
Since $U_0 = \sqrt{2 \pi} g a_z$, $\varepsilon_{dd}$
is also equal to $(\sqrt{8 \pi}/3) D^2/(g a_z)$. This
quantity gives roughly the ratio between the expectation 
value of the dipolar energy 
$E_{\rm dip}$ and $E_s$. Therefore, with the above 
choice of parameters, and for the specific values 
of $D$ that we choose below, we have the following 
hierarchy of energy scales:
\begin{equation}
\hbar \omega_z > E_s \gtrsim E_{\rm dip} > \hbar \omega_0 \;.
\end{equation}
It should be noted that the stability of the system 
is already a non-trivial question \cite{UF,JB}, as the 
dipole-dipole interaction is partly attractive and one 
has to be cautious, since the system may be unstable 
against collapse. We have observed this collapse in
our calculations, although not in the range of 
parameters corresponding to the figures shown below. 

\subsection{Vortex configurations as a function of the dipole 
strength and the dipole orientation}

Let us first study the rotational properties of the system as 
a function of the orientation of the dipoles of the atoms with 
respect to their plane of motion, and of the dipolar strength. 
In Fig.\,2 we show the two-dimensional atom density for 
$\Theta=0$, 15, 30, 45, and 90 degrees, and $\varepsilon_{dd} 
= 0$, 0.31, 0.38, 0.54, and 0.67. We also choose a moderate 
value of $A=-0.03$, which corresponds to a trapping potential 
that is more tight along the $y$ axis and finally $\Omega/
\omega_0$ is set equal to 3/4.
\begin{figure}[t]
\centerline{\includegraphics[width=9cm,clip]{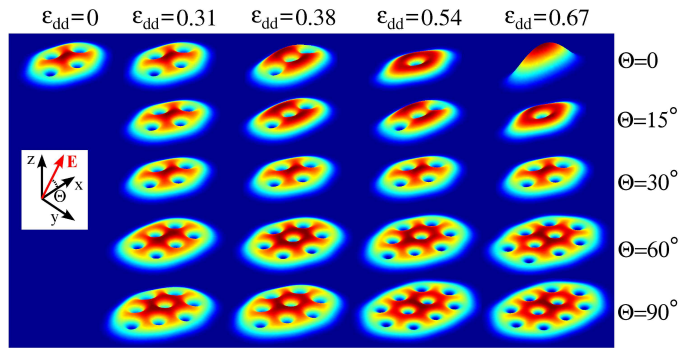}}
\caption{(Color online)
Two-dimensional atom density as a function of the strength 
of the dipolar interaction, for $\varepsilon_{dd} = 0$, 0.31, 
0.38, 0.54, and 0.67, and of the orientation of the 
dipoles, for $\Theta=0$, 15, 30, 60, and 90 degrees, with 
the non-dipolar case ($\varepsilon_{dd} = 0$) also shown 
as a reference. The trap deformation is $A = - 0.03$, and 
the angular velocity of the trap is $\Omega/\omega_0 = 3/4$.
The inset illustrates the coordinate axes and the external field.}
\label{fig2}
\end{figure}
From Fig.\,2 one can see that the way the vortex structure 
changes as a function of $\varepsilon_{dd}$ depends sensitively 
on the orientation of the dipoles. For example, when the 
polarization is along the $x$ axis ($\Theta=0^\circ$) the attractive 
part of the dipolar interaction gives rise to a well-known 
self-induced squeezing of the density in the $y$ direction 
\cite{Abad09} that reduces progressively the number of 
vortices as $\varepsilon_{dd}$ increases.
 
Thus, the four vortices present in the non-dipolar 
case ($\varepsilon_{dd}=0$) and for $\varepsilon_{dd}=0.31$, 
get reduced to three and to one for $\varepsilon_{dd}=0.38$ 
and 0.54, respectively, with a structure that clearly reflects 
the symmetry imposed by the external field along the $x$ axis. 
Finally, when the dipolar strength exceeds a certain value 
($\varepsilon_{dd} \simeq 0.67$), the width of the gas becomes 
too small to accommodate any vortex state. A similar effect has 
been found in non-dipolar condensates which are confined in 
anisotropic potentials, where there is a critical deformation 
of the trap beyond which all the vortices are expelled from 
the cloud \cite{Fet01,Aft09,Fet10}.

When the polarization angle becomes $\Theta=15^\circ$ we find a 
similar, but less pronounced disappearance of the vortex lattice, 
as in this case the attractive part of the dipolar interaction 
becomes less important. For $\Theta=30^\circ$, however, the repulsive 
and the attractive parts of the dipole interaction become comparable, 
and as a result the vortex structure is not affected by the increase 
in $\varepsilon_{dd}$ within the range that we have considered. 

When the angle exceeds 60$^\circ$ the behavior of the system 
changes. In this case, the repulsive part of the interaction 
becomes dominant, giving rise to a stronger net repulsion. As a 
consequence, the number of vortices increases with increasing 
$\varepsilon_{dd}$. This effect is maximal when the dipoles are 
polarized along the $z$ axis ($\Theta=90^\circ$), in which case 
the dipolar interaction is purely repulsive and isotropic.

\subsection{Vortex configurations as a function of the ellipticity of the trap
  and the dipole orientation}

We have also studied the rotational behavior of the system as a 
function of the trap deformation $A$ and of the angle $\Theta$, 
for a fixed dipolar strength $\varepsilon_{dd}=0.67$ and a fixed
rotational frequency of the trap, $\Omega/\omega_0 = 3/4$, as shown 
in Fig.\,3. In particular, we considered the cases of strong 
and moderate deformation of the trap in the $y$ axis ($A=-0.1$ and 
$A=-0.05$, respectively), the non-deformed case ($A=0$), and of
strong and moderate deformation along the $x$ direction ($A=0.1$ 
and $A=0.05$, respectively). 
\begin{figure}[t]
\centerline{\includegraphics[width=9cm,clip]{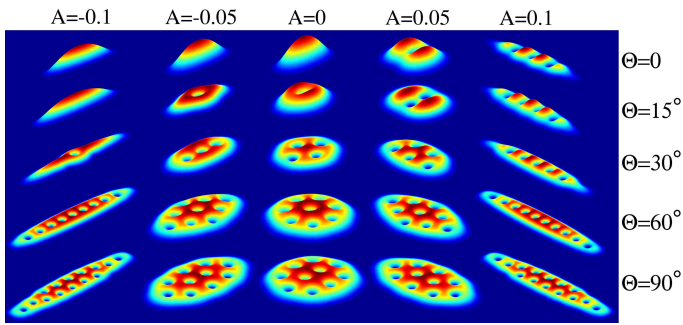}}
\caption{(Color online)
Two-dimensional atom density as a function of the ellipticity 
of the trap for $A=-0.1$, $-0.05$, 0, 0.05 and 0.1, and of the 
orientation of the dipoles, for $\Theta=0$, 15, 30, 60 and 90 
degrees. The angular velocity of the trap is $\Omega/\omega_0 
= 3/4$, and $\varepsilon_{dd}$ is fixed to 0.67. The coordinate
axes correspond to the same as in Fig. 2.}
\label{fig3}
\end{figure}
When the dipoles are polarized in-plane, along the $x$ axis 
($\Theta=0^\circ$), the squeezing of the density in the $y$ 
direction reported in the previous subsection due to 
the attractive part of the dipolar interaction prevents the 
formation of vortices, not only for $A<0$, but also in the 
non-deformed case ($A=0$). However, when the trap is squeezed 
in the direction of the polarization of the dipoles, the 
confinement competes with the dipolar interaction, forcing 
the cloud to expand along the $y$ axis, and thus allowing the 
formation of vortices in the system, with their number 
increasing as the trap gets more deformed.

The same qualitative behavior is found when the polarization
angle $\Theta$ is slightly increased. Indeed, for $\Theta=15^\circ$ 
the main difference with the previous case is that the cloud has 
a vortex state for $A=0$ and $A=-0.05$. Similarly, when the angle 
becomes 30 degrees, the tendency of reducing (increasing) 
the number of vortices when the trap is squeezed in the $y$ ($x$) 
axis is still clear, although in this case the vortex lattice does 
not disappear completely, even for the largest deformation that 
we have considered.

This trend is however no longer observed for $\Theta=60^\circ$. 
In this case, the dipolar interaction is mostly repulsive and 
the number of vortices increases with respect to the non-deformed 
case, regardless of the sign of the deformation. When the latter 
is moderate, the vortex structure is the same for $A=\pm 0.05$, 
and differences are only present in the case of strong deformation. 
Finally, when the dipoles are polarized along the vertical direction, 
i.e., when $\Theta=90^\circ$, the dipolar interaction is isotropic 
and purely repulsive and as a consequence the densities are axially 
symmetric for $A = 0$ and have a mirror symmetry with respect the 
interchange of the $x$ and $y$ axes in the deformed cases.

\subsection{Angular momentum versus the rotational frequency 
of the trap}

Finally, we have studied the expectation value of the angular 
momentum of the system in the $z$ direction, $\langle L_z 
\rangle$, as a function of the rotation frequency $\Omega$, for 
different values of $\Theta$ and $A$, and for a fixed strength 
of the dipolar interaction $\varepsilon_{dd}=0.67$, as shown 
in Fig.\,4. 

The different cases shown in Fig.\,4 share some common 
characteristics. First of all, they are consistent with 
the divergence of the angular momentum when $\Omega \to 
{\rm min}\{\omega_x, \omega_y\}$, i.e. when $\Omega \to 
\omega_y = \omega_0 \sqrt{1 - 4 A}$ since $A$ is chosen 
to be positive. Also, as $\Omega/\omega_0$ approaches the 
limiting value $\sqrt{1 - 4 A}$, the curves become more 
smooth, as the number of vortices increases and the system 
becomes more classical, resembling in this limit solid-body 
rotation. We also observe in Fig.\,4 that for some given 
$\Omega$ and $A$, as $\Theta$ decreases, the angular momentum 
of the gas decreases as well. 

In addition, the results in Fig.\,4 are consistent with 
the intuitive expectation that the more the system is 
distorted from axial symmetry (i.e., as $A$ increases, 
or as $\Theta$ decreases), the more it resists in accommodating 
a vortex state \cite{SFetter}, and as a result the critical 
frequency $\Omega_c$ necessary for vortex nucleation increases. 
This effect is most clearly seen in Fig.\,5, which shows the 
critical value of $\Omega$ for the nucleation of the first vortex 
state as a function of $\Theta$. This increase of $\Omega_c$ is 
associated with a distortion of the axial symmetry of the cloud, 
which may carry angular momentum even when it is vortex-free, as 
required by the irrotational nature of the velocity field. 

In most cases shown in Fig.\,4 there is a 
discontinuous increase of $\langle L_z \rangle$ with 
increasing $\Omega$, which is associated with the progressive 
entry of vortex states into the gas, very much as in the case 
of non-dipolar atoms confined in an isotropic trapping 
potential \cite{Rok99}. One interesting exception is 
shown in the top panel of Fig.\,4, where for $\Theta = 0^\circ$ 
(i.e., when the dipolar interaction breaks the axial symmetry 
in the most extreme way) the angular momentum increases 
continuously from zero. In this range of $\Omega/\omega_0$ 
the density of the cloud is elongated along the direction of 
the polarizing field, breaking the axial symmetry of the potential,
and it is also vortex-free. Beyond a critical rotational frequency 
($\Omega/\omega_0 \approx 0.82$), however, the angular momentum 
shows a sudden increase, which is associated with three vortices 
entering the cloud. On the contrary, for the other values 
of $A = 0.03$ and 0.1 that we have considered, there is a 
discontinuous transition to a state with one vortex, as seen 
in the central and bottom panels of Fig.\,4, where $\langle L_z 
\rangle$ jumps from zero to unity.

\begin{figure}[t]
\centerline{\includegraphics[width=8.5cm,angle=0,clip]{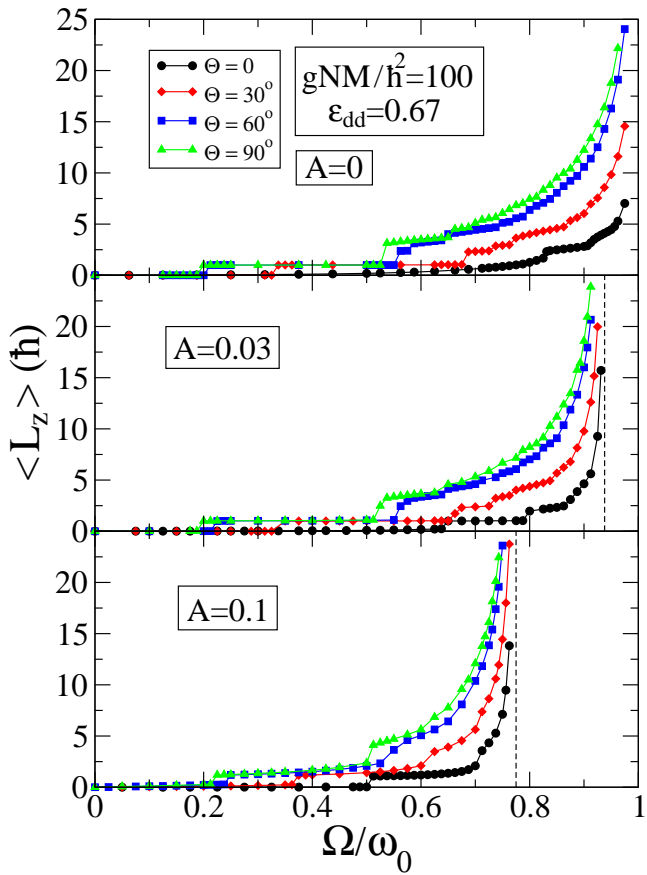}}
\caption{(Color online)
Expectation value of the angular momentum of the gas as a 
function of the rotational frequency of the trap, for $A=0$ (top),
$A=0.03$ (middle) and $A=0.1$ (bottom), for $\varepsilon_{dd}=0.67$, 
and $\Theta=0$, 30, 60, and 90 degrees. The dashed vertical lines
in the lower two curves correspond to the value $\Omega/\omega_0 = 
\sqrt{1 - 4 A}$.}
\label{fig4}
\end{figure}

\begin{figure}[t]
\centerline{\includegraphics[width=9cm,clip]{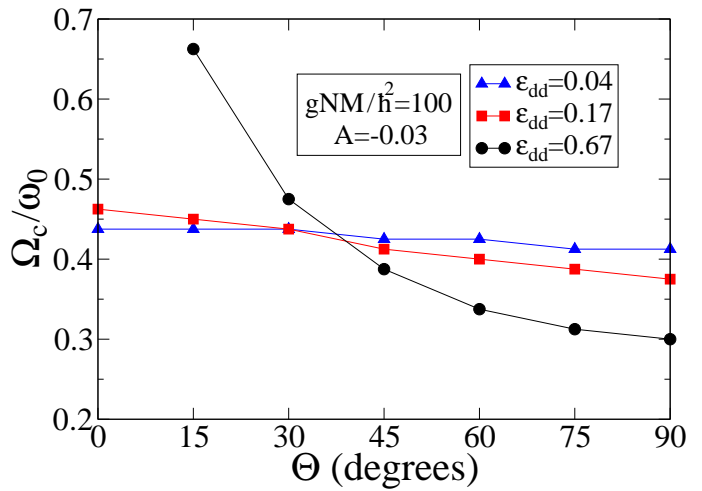}}
\caption{(Color online)
Critical rotational frequency of the trap for the nucleation 
of the first vortex as a function of $\Theta$, for $A=-0.03$, 
and for $\varepsilon_{dd} = 0.04$, 0.17, and 0.67.}
\label{fig5}
\end{figure}

\section{Summary and conclusions}

In this paper we have investigated the stationary states of 
a dipolar Bose-Einstein condensate that is confined in 
a rotating, elliptical, quasi-two-dimensional trapping 
potential. In particular, we examined this problem as a 
function of the orientation of the dipole moment of the 
atoms, of the relative strength between the dipolar and 
the contact interactions, and of the deformation of the 
trap.

While a repulsive contact potential between the atoms tends 
to spread the gas uniformly, the trap deformation and the 
anisotropic character of the dipolar interaction favor the 
breaking of the axial symmetry. These two mechanisms give rise 
to interesting configurations, which we have investigated 
here. For example, at a fixed trap deformation, we observe 
either the disappearance of the vortex lattice, or nucleation 
of vortices as the dipolar strength increases, depending 
on the orientation of the dipoles. A similar behavior 
is found if we fix the strength of the dipole interaction 
and we vary the orientation of the dipole moments, or 
the deformation of the trap.

Another interesting observation is that even in the case
of an axially-symmetric trapping potential, relatively
weak values of the dipolar interaction energy
(as compared to the energy due to the contact potential),
suffice to give rise to rather significant effects due to 
the breaking of the axial symmetry. An analogous 
result has been found in the case of dipole-free atoms
that rotate in an asymmetric trapping potential, where
even weak deviations from axial symmetry have a 
substantial effect \cite{Fet01}. 

One rather general and intuitive observation that 
results from the present study is that the more the 
gas is distorted from axial symmetry (via either the 
deformation of the trap, or via the tilt of the angle
of the dipoles from the direction perpendicularly to 
their plane of motion), the more it expels the vortices, 
with the critical frequency for vortex nucleation 
being increased \cite{SFetter}. 

The strength of the dipolar interaction affects the 
critical frequency for the formation of vortices in 
the cloud in a non-monotonic way. When the dipoles 
are oriented along their plane of motion, or close
to this direction, the rule mentioned in the previous 
paragraph applies and, and as a result, the critical 
rotational frequency for vortex nucleation increases 
with the dipolar interaction strength. On the other 
hand, when the dipoles are either perpendicular to 
their plane of motion, or close to that direction, 
the critical rotational frequency for vortex nucleation 
decreases as the dipolar interaction increases. As 
noticed in Ref.\,\cite{ODell07} this is due to the 
fact that when, for example, $\Theta = 90^\circ$, 
the dipolar interaction is purely repulsive and 
isotropic, and as a result the average atom density 
becomes lower. While the dipolar interaction affects 
the large-scale properties of the cloud (i.e., its radius), 
it also leaves its small-scale properties (i.e., the 
coherence length) unaffected, which is reflected in 
the decrease of $\Omega_c$. 

Our study gives just a flavor of the richness of the 
physical effects that result mainly from the anisotropic 
nature of the dipolar interaction, but also from the 
deformation of the trapping potential. In particular, 
the sensitivity of the behavior of the gas on the 
orientation of the external polarizing field (and 
thus of the dipole moment of the atoms), which may be 
tuned easily, is remarkable. The experimental investigation 
of these effects appears therefore worthwhile.

\section{Acknowledgements}

We thank Niels S{\o}ndegaard for valuable assistance in the
FFT treatment of the dipolar interaction. We also thank 
Manuel Barranco, Georg Bruun, Jonas Cremon, and Mart\'i Pi 
for useful discussions. This work was financed by the Swedish 
Research Council.

\end{document}